\documentclass[superscriptaddress,twocolumn, 10pt]{revtex4}
\usepackage{amsmath,lscape,sidecap,epsfig}
\usepackage{graphicx,color,xcolor}
\usepackage{grffile}
\usepackage{float}
\usepackage{ulem,soul}
\usepackage{lipsum}

\usepackage[charter]{mathdesign}
\DeclareFontFamily{U}{dutchcal}{\skewchar\font=45 }
\DeclareFontShape{U}{dutchcal}{m}{n}{<-> s*[1.0] dutchcal-r}{}
\DeclareFontShape{U}{dutchcal}{b}{n}{<-> s*[1.0] dutchcal-b}{}
\DeclareMathAlphabet{\mathlcal}{U}{dutchcal}{m}{n}
\SetMathAlphabet{\mathlcal}{bold}{U}{dutchcal}{b}{n}

\begin{document}

\title{Fluctuations in the pasta phase} 

\author{Mateus R. Pelicer}
\affiliation{Depto de F\'{\i}sica - CFM - Universidade Federal de Santa
Catarina  Florian\'opolis - SC - CP. 476 - CEP 88.040 - 900 - Brazil}
\author{D\'ebora P. Menezes}
\affiliation{Depto de F\'{\i}sica - CFM - Universidade Federal de Santa
Catarina  Florian\'opolis - SC - CP. 476 - CEP 88.040 - 900 - Brazil}
\author{Celso C. Barros Jr.}
 \affiliation{Depto de F\'{\i}sica - CFM - Universidade Federal de Santa
 Catarina  Florian\'opolis - SC - CP. 476 - CEP 88.040 - 900 - Brazil}
\author{Francesca Gulminelli}
\affiliation{CNRS and ENSICAEN, UMR6534, LPC, \\ 
14050 Caen c\'edex, France}

\begin{abstract}
Baryonic matter close to the saturation density is very likely to present complex inhomogeneous structures collectively known under the name of pasta phase. At finite temperature, the different geometric structures are expected to coexist, with potential consequences on the neutron star crust conductivity and neutrino transport in supernova matter. In the framework of a statistical multi-component approach, we calculate the composition of matter in the pasta phase considering density, proton fraction, and geometry fluctuations. Using a realistic energy functional from relativistic mean field theory and a temperature and isospin dependent surface tension fitted from Thomas-Fermi calculations, we show that different geometries can coexist in a large fraction of the pasta phase, down to temperatures of the order of the crystallization temperature of the neutron star crust. Quantitative estimates of the charge fluctuations are given.

\end{abstract}
\maketitle

{\it Introduction:}
It is known since the early '80s \cite{PhysRevLett.50.2066} that the equilibrium state of electrically neutral dense baryonic matter, close to the saturation density ($n_{sat}\approx 0.17$ fm$^{-3}$) of symmetric nuclear matter, may not correspond to a lattice of spherical nuclei, but rather to a spatially periodic distribution of inhomogeneities with cylindrical (rods) or planar (slabs) symmetry. These complex pasta phases could be present in the inner crust of neutron stars as well as in the central regions of core collapse supernovae during the infall and early post-bounce phase. Different studies suggest that they may have sizeable impact on different astrophysical phenomena, such as the magneto-thermal evolution of compact stars \cite{Schmitt2018}, neutrino opacity \cite{PhysRevC.70.065806,2016arXiv161110226H}, timing properties of X-ray pulsars \cite{Pons2013}, and the ellipticity of neutron stars that can be potentially probed by gravitational wave measurements \cite{Gearheart2011}. The energy barrier between the different shapes being often of a few KeV only, it is expected that complex shapes including impurities and defects might appear even in the crust of catalyzed neutron stars \cite{Schneider2016,PhysRevC.98.055801,Nandi_2018} leading to an increased resistivity. Above the crystallization temperature, $T_m\approx 700$ KeV in the pasta region \cite{Carreau:2019tiv}, pasta matter has to be seen as a liquid and the calculation of transport properties requires the evaluation of the electrical and thermal conductivity tensor averaged over finite domains. A rich phenomenology is expected depending on the orientation of the nuclear clusters \cite{Yakovlev2016} as well as their distribution. 

In this paper, we make a first step towards the complex problem related to the anisotropic transport in the pasta phase at finite temperature, by calculating the distribution and charge variance of the different geometrical structures in a multi-component approach with a realistic microscopic energy functional.

{\it Pasta phases in a Relativistic Mean Field Approach:} 
 
In the single nucleus approximation it is assumed that, at a given thermodynamical condition $(\rho_B, Y_p, T)$, a crystalline structure of identical cells is formed. Each cell contains a dense cluster at baryonic density $\rho^I=\rho_p^I+\rho_n^I$,  occupying a volume fraction $f$ of the cell, and surrounded by a homogeneous gas of baryons at density $\rho^{II}=\rho_p^{II}+\rho_n^{II}$. The cell is neutralized by an electron gas of homogeneous density $\rho_e= Y_p \rho_B$, modelled as a free relativistic gas with free energy density ${\cal F}_e$, see Refs.~\cite{PhysRevC.78.015802, PhysRevC.79.035804} for explicit expressions. 
The different shapes of the pasta structures is denoted by the integer $d=3, 2, 1$, referring to  droplets, rods and slabs, respectively.  

The free energy density of a cell of total baryonic density $\rho_B = \rho_p + \rho_n$ and proton fraction $Y_p=\rho_p/\rho_B$ is given by~\cite{PhysRevC.78.015802}
\begin{flalign}\label{eq:F_SNA}
    {\cal F}_{WS}= & f {\cal F}_b^I+(1-f){\cal F}_b^{II}+ \beta {\cal F}_{sc, d} + {\cal F}_e,
\end{flalign}
with $\beta=f$ for droplets, rods and slabs, and $\beta=1-f$ for tubes and bubbles. The constraints of  mass and charge conservation are imposed on the cells, $\rho_q=f (\rho_{q}^I - \rho_{q}^{II}) + \rho_{q}^{II}$, with $q=n,p$.

For all our numerical applications, the bulk free energy densities ${\cal F}_b^{I(II)}={\cal F}_b(\rho_{p }^{I(II)},\rho_{n}^{I(II)} )$  
of the dense (dilute) fraction of the cell are calculated with
the IUFSU parametrization~\cite{IUFSU} of the 
quantum hadrodynamics model 
in the mean field approximation (RMF)~\cite{WALECKA1974491}
which is consistent with a number of experimental and observational constraints~\cite{PhysRevC.90.055203, PhysRevC.93.025806,PhysRevC.99.045202}. This version of the RMF model
includes $\sigma, \omega$ and $\rho$ mesons with non-linear scalar and vector couplings, as well as an $\omega-\rho$ interaction term, with a coupling fitted such as to reproduce the experimental symmetry energy of nuclear matter~\cite{IUFSU}.

The interface free energy density  ${\cal F}_{sc,d}$ contains a surface and a Coulomb term and it is written as~\cite{PhysRevC.72.015802, PhysRevLett.50.2066}:
 \begin{equation}\label{eq:inmedium}
    {\cal F}_{sc, d}= {\cal F}_{s, d}+ {\cal F}_{c, d}=\frac{\sigma(Y_p, T) \, d }{R_d} + 2 \pi e^2 R_d^2 \left( \rho_{p }^I- \rho_{p }^{II} \right)^2 \Phi_d(\beta),
 \end{equation}
 with the $\Phi_d$ function given by
\begin{equation} \label{eq:phid}
\Phi_d(\beta)=\begin{cases}
\left(\frac{2-d \beta^{1-2/d}}{d-2}+\beta \right) \frac{1}{d+2}, \quad d=1,3;\\
 \frac{\beta-1-ln(\beta)}{d+2}, \quad d=2. \end{cases}
\end{equation}
The temperature and proton fraction dependent surface tension coefficient $\sigma$ is taken from Ref. \cite{PRC85}, where a fit was obtained from 
Thomas-Fermi calculations employing the same IUFSU functional.

The equilibrium densities in the cells can be found by minimizing the thermodynamic potential $\Omega$ 
with respect to the 6 independent variables, here chosen as the cluster and gas densities $\rho_{q}^I,  \rho_{q}^{II}$, $q=n,p$, the volume fraction $f$, and the linear dimension of the pasta $R_d$:
\begin{equation}\label{eq:auxiliary}
    {\Omega} = {\cal F} - \mu_p \rho_p - \mu_n \rho_n,
\end{equation}
with the conservation laws imposed by the chemical potentials, $\mu_p$ and $\mu_n$.

The equilibrium equations from the minimization of Eq.~\eqref{eq:auxiliary} are obtained as:
\begin{flalign}\label{eq:Lp1}
\mu_{p}^I = \mu_{p}^{II}  - \frac{2 \beta {\cal F}_{c,d}}{f(1-f)(\rho_{p}^I - \rho_{ p}^{II})} ,
\end{flalign}
\begin{flalign}\label{eq:Ln1}
\mu_{n}^I =\mu_{n}^{II}, 
\end{flalign}
\begin{flalign}
- P^I +&P^{II} + \frac{d \beta}{d f}\left(  {\cal F}_{sc,d} + \beta {\cal F}_{c,d} \, \frac{1}{\Phi} \frac{d \Phi}{d \beta}  \right) \nonumber\\
&- \frac{2 \beta {\cal F}_c\left( \rho_{p}^I - f(\rho_{p}^I - \rho_{p}^{II}) \right)}{f (1-f) \left(\rho_{p}^{II} - \rho_{ p}^{II}\right)} = 0, \label{eq:Lf1}
\end{flalign}
\begin{flalign}
R_d=\left( \frac{\sigma d}{4 \pi e^2 (\rho_{p}^I-\rho_{p}^{II})^2 \Phi_d (\beta)} \right)^{1/3},
\label{eq:LR1}
\end{flalign}
where $\mu_q^K, P^K$,q=n,p, K=I,II represent bulk chemical potentials and pressures, $\mu_q^K=\partial {\mathcal F}_b^K/\partial \rho_q^K$, $P^K=-{\cal F}_b^K+\sum_{q}\mu_q^K\rho_q^K$ .
Eqs.(\ref{eq:Lp1})-(\ref{eq:LR1}) are in agreement with those of Refs.~\cite{Pais:2018ilv, PhysRevC.91.055801}.
The variation of the thermodynamic potential additionnally allows us to determine the thermodynamic chemical potentials for the inhomogeneous system, given as:
\begin{equation}\label{eq:Lambda}
   \mu_q = \mu_{q}^{I} + \frac{\beta}{f} \frac{\partial {\cal F}_{sc, d}}{\partial \rho_{q}^{I}} = \mu_{q}^{II} + \frac{\beta}{1-f} \frac{\partial {\cal F}_{sc, d}}{\partial \rho_{q}^{II}}.
\end{equation}
The minimization is done independently for the different geometries, and the one corresponding to the minimum value of the optimized
${\cal F}_{WS}$ is associated to the equilibrium configuration.

{\it Fluctuations: distribution of pasta structures}

The SNA is not realistic in the sense that different cluster sizes and geometries can coexist in a macroscopic system due to the small difference in free energy densities between them. Following Refs.~\cite{Gulminelli_2015,PhysRevC.97.035807,Carreau:2020gth,PhysRevC.101.035211}, we   consider a macroscopic volume $V$ composed of different Wigner-Seitz cells of volume $V_{WS}^N$. Each cell is composed of unbound nucleons and electrons gases, which we assume to be of constant density over the entire volume to avoid discontinuities in the chemical potential between cells. Since the SNA is known to provide a good description of the average thermodynamic quantities of the system, the densities of the nucleon gas $\rho_{p g}, \rho_{n g}$ are given by the solution of the coupled equations 
(\ref{eq:Lp1})-(\ref{eq:LR1}), $\rho_{q,g}=\rho_q^{II}$, and the electron density is $\rho_e=\rho_p$.
The center of each cell is occupied by a cluster in the pasta phase, with 
proton and neutron densities $\rho_p^N,\rho_n^N$ fluctuating from cell to cell, and we introduce a superscript $N$ to all the variables which vary with the density fluctuations. 
We consider that for a same total density $\rho_B$, that is at a given depth inside the star, domains with pasta structures corresponding to different geometries ($d=1, 2, 3$ for slabs, rods and droplets, respectively) may also coexist, due to the small free energy barriers. 

In the presence of clusters with fluctuating densities and shapes, the global densities are given by
\begin{equation}\label{eq:globalProton}
    \rho_q =  \sum_{N,d} n^{N,d} V^N \left( \rho_{q }^N - \rho_{q g} \right) + \rho_{q g},  
\end{equation} 
where $V^N=f^N V_{WS}^N$ is the volume of the cluster, and $n^{N,d}= {\cal N}^{N, d}/V$ is the number density of a cell containing a cluster of density $\rho^N=\rho_p^N+\rho_n^N$ and dimension $d$, normalized to  $\sum_{N, d} n^{N, d} V_{WS}^N =1$.

In principle, both the cluster volume $V^N$ and the volume fraction $f^N$ could fluctuate independently of $\rho_p^N$,$\rho_n^N$ and $d$. However, in a variational theory the linear size of the cluster is determined by the equilibrium with the surrounding gas, whose density is considered constant throughout the system. We therefore consider that the minimization with respect to the linear size Eq.(\ref{eq:LR1}) holds in each cell, yielding: 
\begin{flalign}
R_d^N=\left( \frac{\sigma d}{4 \pi e^2 (\rho_{p}^N-\rho_{p g})^2 \Phi_d (\beta^N)} \right)^{1/3}.
\label{eq:LRN}
\end{flalign}
 
Nevertheless, we consider the cluster volume to be independent of its geometry. For a given fluctuation ($\rho_p^N,\rho_n^N$), the reference volume corresponds to the spherical one
\begin{equation}\label{eq:volume}
V^N=4\pi \left (R_3^N\right ) ^3/3,
\end{equation}
and the Wigner-Seitz cell volume is $V_{WS}^N = V^N /f^N$. The number of protons in the cluster will therefore be given by:
\begin{equation}
    Z^N= \left(\rho^N - \rho_{p g} \right) V^N,
\end{equation}
independently of $d$.
Moreover, we neglect possible long range Coulomb interactions between neighboring cells, by imposing charge neutrality in each cell, thus fixing the cluster volume fraction $f^N$:
\begin{equation}
    \rho_e =f^N (\rho_{p }^N - \rho_{p g}) + \rho_{p g}.\label{eq:rhopcell} \\
\end{equation}

Within these simplifications, we have only two independent variables that can fluctuate from cell to cell, that we take to be ($\rho_p^N,\rho_n^N$). The free energy density of the global system can be written as
$ {\cal F}= {\cal F}_{cl} +{\cal F}_g +  {\cal F}_e$,
with ${\cal F}_g={\cal F}_b^{II}$, as obtained from the SNA calculation, and the cluster component corresponding to a linear-mixing multi-component plasma expression that reads:
 
\begin{equation}\label{eq:F_NSE}
    {\cal F}_{cl}=\sum_{N,d} n^{N, d} V^N \left ( {\cal F}_b^N -{\cal F}_g +{\cal F}_{sc, d}^N \right ) .
\end{equation}

Because of the additivity of the free energy densities, the grand-partition function of the macroscopic system can be factorized in terms comprising the gases and the clusters ~\cite{Gulminelli_2015}:
\begin{equation}\label{eq:partitionFac}
    {\mathlcal Z} = {\mathlcal z}_g^V {\mathlcal z}_e^V  {\mathlcal Z}_{cl} 
\end{equation}
with each term being given by
\begin{flalign}
{\mathlcal z}_e &= \exp\left[-\beta \left({\cal F}_e - \mu_e \rho_e\right) \right] ,\\
{\mathlcal z}_g &= \exp\left[-\beta \left({\cal F}_g -  \mu_n \rho_{n g} - \mu_p \rho_{p g}\right) \right] ,\\ 
{\mathlcal Z}_{cl} &= \sum_{ \{ n\} } \exp\left[ -\beta V \sum_{N, d} n^{N, d} \tilde \Omega^{N, d} \right]. \label{eq:zclus}
\end{flalign}

Here, the chemical potentials are taken from Eq.(\ref{eq:Lambda}), and the first sum in Eq.(\ref{eq:zclus})
is extended to all the possible values of $n^{N, d}$.
$\tilde \Omega_{WS}^{N, d}$  is the single-cluster grand-canonical potential, given by~\cite{Gulminelli_2015}:
\begin{flalign}\label{eq:omega}
    \tilde\Omega^{N, d}= V^N  
    & \left[\frac{\partial  {\cal F}_{cl}}{\partial n^{N, d} }- \sum_q \mu_q\left( \rho_{q}^N - \rho_{q g} \right)  \right ].
\end{flalign}

In calculating the variation $\partial {\cal F}_{cl}/\partial n^{N, d}$ we can see that a rearrangement term appears:
\begin{flalign} \label{eq:FNdef}
    \frac{\partial  {\cal F}_{cl}}{\partial n^{N, d} } =   &  F^{N,d} + \sum_{M,d'} n^{M,d'} \frac{\partial F^{M,d'}}{\partial n^{N,d}},
\end{flalign}
with $F^{N,d}=V^N \left[ {\cal F}_b^N - {\cal F}_g + {\cal F}_{sc, d}^N\right]$ being the $N$-cluster free energy. 
Rearrangement terms systematically appear in thermodynamics when the appropriate thermodynamic potential (here: the Helmotz free energy density ${\cal F}_{cl}$)  depends on (one of) the system densities. Indeed a variation of a specific cluster density $n^{N, d}$ according to Eq.(\ref{eq:FNdef}) induces a variation of the global densities $\rho_q$, as it can be seen from  Eq.(\ref{eq:globalProton}), and consequently of the thermodynamic potential. In our model, $F^{N,d}$ does not specifically depend on $\rho_n$ and $\rho_p$, but a dependence on $\rho_p$ is enforced by the neutrality condition inside the cell Eq.(\ref{eq:rhopcell}), that can be viewed, for a given fluctuation $(\rho_p^N,\rho_n^N)$, as a definition of the cell volume fraction as a function of the electron density, $f^N\equiv f^N(\rho_p)$. 

  Using Eqs.~\eqref{eq:globalProton} and~\eqref{eq:rhopcell} we immediately get:
  \begin{flalign}
   \frac{\partial  {\cal F}_{cl}}{\partial n^{N, d} } &=     F^{N,d} +
        \frac{\partial \rho_p}{\partial n^{N,d}}\sum_{M,d'} n^{M,d'} \frac{\partial F^{M,d'}}{\partial f^M} \frac{\partial f^M}{\partial \rho_p}    \label{eq:rear} \\
        &=  F^{N,d} +
         V^N\left( \rho_{p }^N - \rho_{p g} \right) \sum_{M,d'} n^{M,d'} \frac{\partial F^{M,d'}}{\partial f^M} \frac{V^M}{\rho_{p }^M - \rho_{p g} } . \nonumber
 \end{flalign}
 If we additionally assume that the different averaged quantities are not correlated, such that the average commutes with the product, we end up with the same expression proposed in Refs.\cite{PhysRevC.97.035807,Carreau:2020gth,PhysRevC.101.035211} :
 \begin{flalign}
      \frac{\partial  {\cal F}_{cl}}{\partial n^{N, d} } &=     F^{N,d} +  
       V^N\left( \rho_{p }^N - \rho_{p g} \right) \left \langle \frac{f^M}{\rho_{p }^M - \rho_{p g}} \frac{\partial {\cal F}_{sc, d}^M}{\partial f^M}  \right\rangle ,
  \end{flalign}
where the notation $\langle X \rangle$ indicates an ensemble average, and can be identified with the value of the $X$ variable taken in the SNA approximation with the optimal geometry $d_0$:
\begin{flalign}
&   \left \langle \frac{f^M}{\rho_{p }^M - \rho_{p g}} \frac{\partial {\cal F}_{sc, d}^M}{\partial f^M}  \right\rangle
   = \frac{ {\cal F}_{c,d_0}}{\Phi}\frac{d\Phi}{df}\nonumber \\ &+ \frac{d_0 f}{R_{d_0} \rho_B}\frac{\left(\rho_{p}^I - \rho_{p g}\right) - Y_p \left( \rho^I - \rho_g\right) }{\rho_{p }^I - \rho_{p g}} \frac{\partial \sigma}{\partial Y_p} .
\end{flalign}

Since the number of occurrences of the different configurations ${\cal N}^{N, d}$ in the thermodynamic limit can be any integer $m\ge 0$, the sum in Eq.(\ref{eq:zclus}) can be analytically computed as in Ref.~\cite{PhysRevC.97.035807}:
\begin{equation} \label{eq:partition_cl}
    {\cal Z}_{cl}= \prod_{N, d}\sum_{m=0}^\infty \frac{\left ( \exp \left[-\beta \tilde \Omega^{N, d} \right ]\right )^m}{m!}
    = \prod_{N, d} \exp \omega^{N, d} ,
\end{equation}
with $\omega^{N, d}=\exp\left(-\beta \tilde \Omega^{N, d}\right)$.

We can remark that the cluster partition sum Eq.(\ref{eq:partition_cl}) has the same functional form as the ideal gas, with a reduced density $\rho_{q}^N-\rho_{q g}$ in order to fulfill particle number conservation. The equilibrium chemical potential of the clusters can be immediately deduced: %
\begin{equation}
  \mu^N=f^N V_{WS}^N\left ( \mu_n (\rho_{n}^N-\rho_{n g})+  \mu_p (\rho_{p}^N-\rho_{p g}) \right ) .
\end{equation} 

  The equilibrium number density for a fluctuation $(\rho_p^N,\rho_n^N)$ is then readily found from the cluster partition sum, and by summing up  the different geometries:  
\begin{equation}
n^{N}=\frac{1}{V}\frac{\partial \ln {\mathlcal Z}_{cl}}{\partial \beta\mu^N}= \sum_{d=1}^3 n^{N,d} = \sum_{d=1}^3 \frac{\omega^{N, d}}{V}
\end{equation}
 
Finally, the probability of a cluster with density $(\rho_p^N,\rho_n^N)$ and geometry $d$ is:
\begin{equation} \label{eq:proba}
    p^{N, d} = \frac{n^{N, d}}{\displaystyle\sum_{N, d} n^{N, d}} = \frac{\exp{ \left(-\beta \tilde \Omega^{N, d} \right) } }{\displaystyle \sum_{N, d} \exp{\left(-\beta \tilde  \Omega^{N, d}\right) }}.
\end{equation}

\begin{figure*}
    \centering
    \includegraphics[scale=0.8]{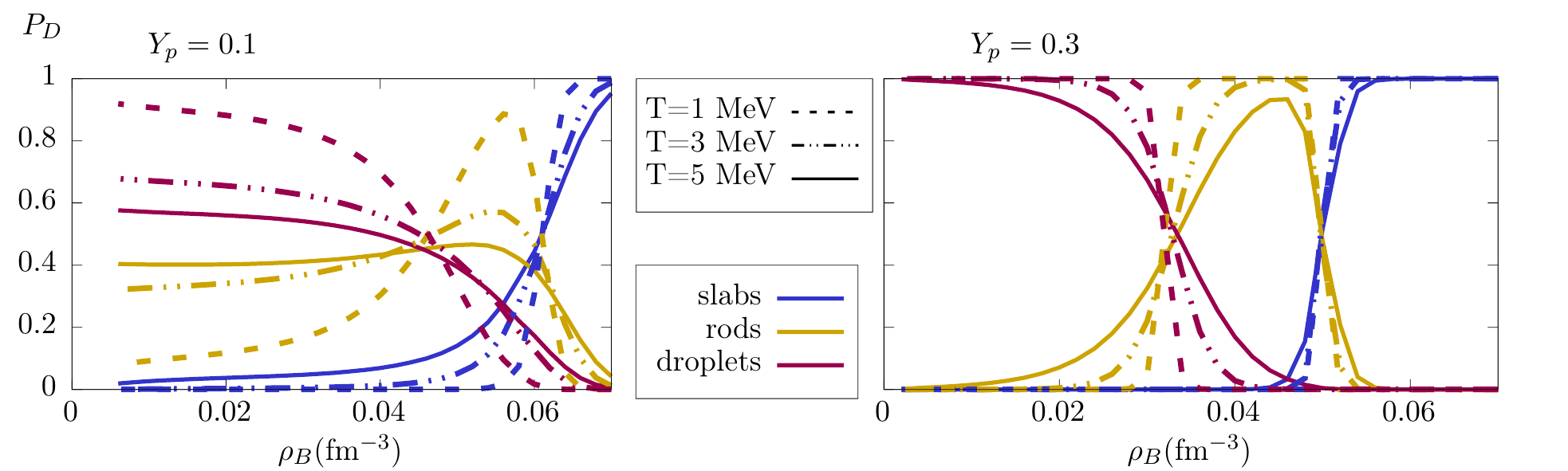}
    \caption{Probability for the pasta structures at proton fractions $Y_p=0.1$ (left) and $0.3$ (right) and temperatures T=5, 3, 1 MeV plotted with solid, dash-double-dotted and dashed lines, respectively. Geometries are plotted with different colors: droplets are in magenta, rods in yellow and slabs in blue. }
    \label{fig:probpasta03}
\end{figure*}

{\it Results:}

To illustrate the proportion of different pasta structures at different thermodynamic conditions we show, in Fig.~\ref{fig:probpasta03}, the probability of each pasta structure for $Y_p=0.1$ and $0.3$, which are  typical proton fractions encountered in  the neutron star inner crust, and in supernova cores, respectively.     We define the probability of a pasta structure by summing the probabilities of its occurence with different densities, $P^d=\sum_N p^{N,d}$.
For increasing temperatures or isospin asymmetry, different pasta structures are more likely to coexist. However, shape coexistence cannot be neglected in a large range of densities even for the typical thermodynamic conditions of the inner crust at the crystallization point ($T\approx 1$ MeV, $Y_p\approx 0.1$ \cite{Carreau:2019tiv}), suggesting that important impurities may characterize even the catalyzed crust\cite{Nandi_2018,PhysRevC.98.055801}.

\begin{figure}[!b]
    \centering
    \includegraphics[scale=0.8]{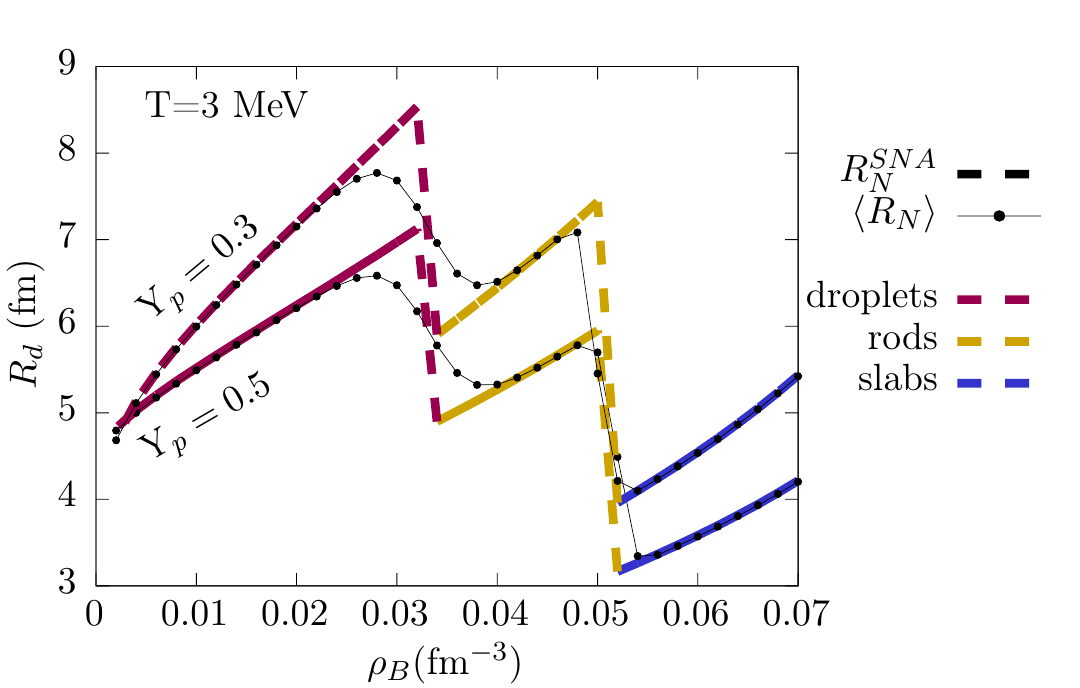}
    \caption{Average linear radius (dashed-dotted) and equilibrium linear radius (colored dashed), as given by Eqs.~\eqref{eq:avgRN}  and~\eqref{eq:LR1}, for proton fractions $Y_p=0.1$, 0.3 and T=3 MeV. The values in the SNA are colorred according to the geometry, as described in Fig.~\ref{fig:probpasta03} .}
    \label{fig:radius1}
\end{figure}

The linear dimension of the structures is measured by the average radius
\begin{equation}\label{eq:avgRN}
    \langle R^N \rangle = \sum_{N, d} p^{N,d} R_d^N,
\end{equation}
illustrated for $T=3$ MeV and $Y_p=0.3$ and 0.5 in Fig.~\ref{fig:radius1}, where the dominant pasta structure occupies most of the system volume fraction away from the transition density between geometries. In such regions the average radius and the one obtained from the SNA solution -- Eq.~\eqref{eq:LR1} -- coincide, but differences are observed in the transition regions.

\begin{figure*}
    \centering
    \includegraphics[scale=0.7]{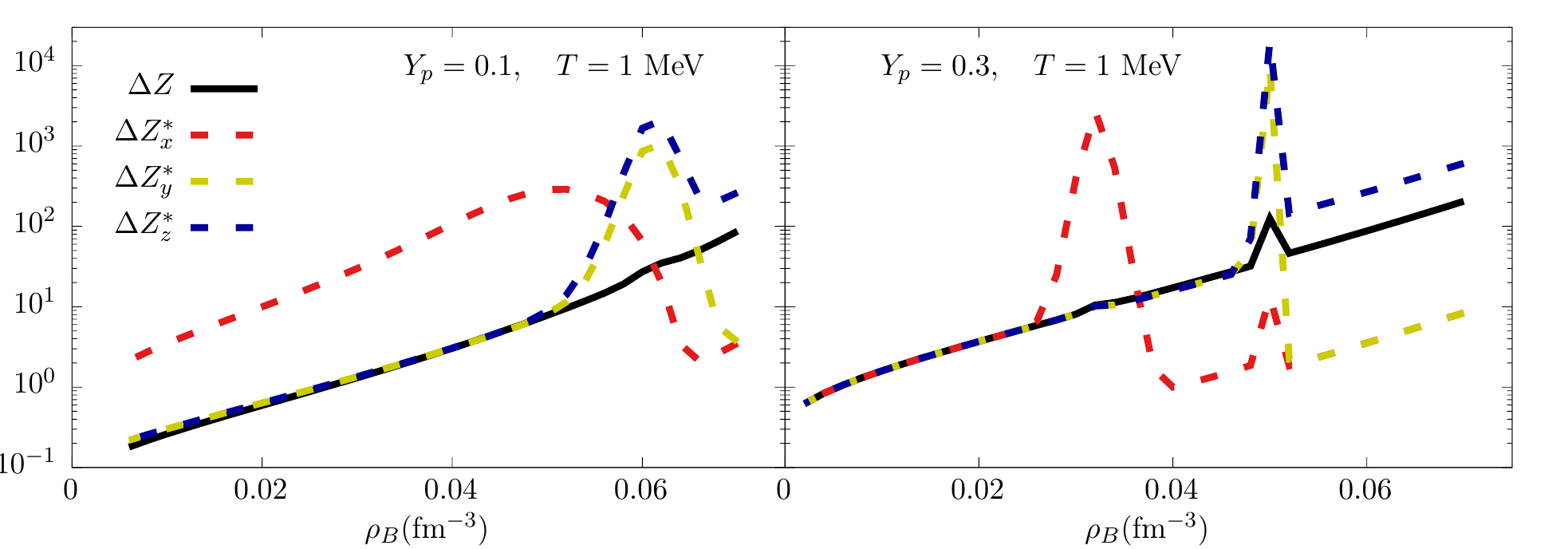}
    \caption{Total  (solid line) and orientation-dependent (dashed) proton number variances for T=1 MeV,  $Y_p=0.1$ (left) and 0.3 (right). The x, y and z axis denote the direction of motion of the probe, and are represented by the colors red, yellow and blue, respectively. We assume local domains of different geometries aligned along a common symmetry axis $x$. }
    \label{fig:impurity2}
\end{figure*}

Finally, the disorder of the composition can be quantified by the variance of proton number:
\begin{equation}
\Delta Z = \sum_{N,d} p^{N,d}\left( Z^N - \langle Z^N \rangle \right)^2, \label{eq:variance}
\end{equation}
that is displayed in Fig.\ref{fig:impurity2} for T=1 MeV, as a solid line. As observed in previous works \cite{Carreau:2020gth,PhysRevC.101.035211}, this variance is an increasing function of the baryonic density and temperature (not shown).
Notice that the variance is mostly independent of the presence of different geometries due to the 
normalization to the size of the spherical cell.


To quantify the impurities present due to the different pasta structures, we must consider that an incident probe is sensitive to their orientation \cite{Yakovlev2016}. A full calculation of the transport coefficients in the different thermodynamic conditions, and according to the possible pasta orientations is left for future work. However, we can appreciate the importance of the anisotropies due to the impurities, by computing the effective charge variance in the hypothesis of a common symmetry axis $x$ for the different local domains characterized by a given geometry.
To do so, we consider the free parameters $L_{d, WS}$ in the volume of slabs and rods:
\begin{equation}
    V^N = \pi R_2^N L_{2, WS}, \quad \textrm{and} \quad
    V^N= R_1^N L_{1, WS}^2
\end{equation}
which can be determined by equating the above to the cluster volume defined in Eq.~\eqref{eq:volume}, in order to define the effective number of protons as seen by an incident probe
\begin{equation}
    Z^{*\, N}_{d, k} = Z^N S_{d, k}^N/S_3^N,
\end{equation}
where k=x, y, z is the axis of motion of the incident probe, $S_{d, k}$ is the surface area of the pasta geometry perceived by the probe, summarized in Tab.~\ref{tab:surfaces} and $S_3^N = 2\pi (R_3^N)^2$ is half of the droplet total surface.  
\begin{table}[!h]
    \centering
    \setlength{\tabcolsep}{10pt}
\renewcommand{\arraystretch}{1.3}
    \begin{tabular}{|c|c|c|}    \hline
        Axis & $S_{2, k}$ & $S_{1, k}$\\ \hline
         x & $\pi (R_2^N)^2$ & $R_1^N L_{1, WS}$ \\ \hline
         y & $\pi R_2^N L_{2, WS} $ & $R_1^N L_{1, WS}$ \\ \hline
         z & $\pi R_2^N L_{2, WS} $ &  $L_{1, WS}^2$\\ \hline
    \end{tabular}    
    \caption{ Effective surface of the pasta structure in the axis of motion of an incident probe. The rods are considered to have length $L_{2, WS}$ in the x-axis, and the slabs have length $L_{1,WS}$ in the x and y directions.}
    \label{tab:surfaces}
\end{table}
Thus we can calculate the variances for the different orientations using Eq.(\ref{eq:variance})
by replacing $Z^N\rightarrow Z^{\ast N}_{d, k}$.
The orientation-dependent variances are also shown in Fig.~\ref{fig:impurity2}. Though the order of magnitude of this effective impurity factor is consistent with previous estimations \cite{Pons2013,Schneider2016}, we can see that huge fluctuations are expected as a function of the orientation, at the densities corresponding to a change of dominant geometry.

{\it Conclusions:}
Using a realistic microscopic energy functional (the IUFSU parametrization with Thomas-Fermi fitting for the surface energy),  we calculated the probability of different pasta structures coexisting at thermodynamic conditions experienced in neutron star and supernovae matter, and made estimates of the proton number variance, considering the different orientations of the pasta structures. Quantitative calculation of the transport coefficients are in progress.

\newpage

{\bf Acknowledgments} 
This  work is a part of the project INCT-FNA Proc. No. 464898/2014-5. D.P.M. is  partially supported by Conselho Nacional de Desenvolvimento Cient\'ifico  e  Tecnol\'ogico  (CNPq/Brazil) under grant  301155.2017-8  and  M.R.P. is supported  with  a  doctorate  scholarship by Coordena\c c\~ao de Aperfei\c coamento de Pessoal de N\'ivel Superior (Capes/Brazil).

\bibliography{paper0105}

\end{document}